# Development and Evaluation of Adaptive Learning Support System Based on Ontology of Multiple Programming Languages

Lalita Na Nongkhai
Graduate School of Engineering, Kochi University of Technology, Japan
256010u@gs.kochi-tech.ac.jp

Jingyun Wang
Department of Computer Science, Durham University, UK
jingyun.wang@durham.ac.uk

Takahiko Mendori
Graduate School of Engineering, Kochi University of Technology, Japan
mendori.takahiko@kochi-tech.ac.jp

**Abstract**

This paper introduces an ontology-based approach within an adaptive learning support system for computer programming. This system (named ADVENTURE) is designed to deliver personalized programming exercises that are tailored to individual learners' skill levels. ADVENTURE utilizes an ontology, named CONTINUOUS, which encompasses common concepts across multiple programming languages. The system leverages this ontology not only to visualize programming concepts but also to provide hints during practice programming exercises and recommend subsequent programming concepts. The adaptive mechanism is driven by the Elo Rating System, applied in an educational context to dynamically estimate the most appropriate exercise difficulty for each learner. An experimental study compared two instructional modes, adaptive and random, based on six features derived from 1,186 code submissions across all the experimental groups. The results indicate significant differences in four of six analyzed features between these two modes. Notably, the adaptive mode demonstrates a significant difference over the random mode in two features: the submission of correct answers and the number of pass concepts. Therefore, these results underscore that this adaptive learning support system may support learners in practicing programming exercises.

## 1. Introduction

Personalized learning is an approach that focuses on individual learners' needs and goals (Shemshack et al., 2021; Taylor et al., 2021). It involves various elements, including individualized strategies, content, activities, and real-time feedback (Shem-shack & Spector, 2020). However, designing personalized instruction in traditional classroom settings for many learners can be challenging. As a result, recent studies emphasize delivering individualized learning experiences through digital platforms.

One promising technological innovation for personalized learning is adaptive learning (Burbules et al., 2020). "Adaptive learning technology provides personalized learning at scale by accessing learners' current skills/knowledge" (Taylor et al., 2021). It can utilize real-time data to automatically adjust the learning experience for individuals based on their performance and behaviors (Anindyaputri et al., 2020; Martin et al., 2020). Researchers have applied adaptive learning to improve specific skills in learners by providing tailored tasks and exercises at dynamic difficulty levels (Fabic et al., 2018; Gavrilović et al., 2018; Troussas et al., 2021).

To facilitate the implementation of adaptive learning techniques, a learning support system may provide environments built on various structures, frameworks, or platforms designed to enhance the educational process by offering resources, tools, and services to learners (Zuhairı et al., 2007). Numerous studies have integrated learning support systems with different learning theories and methodologies, such as collaborative learning support systems (Jeong & Hmelo-Silver, 2016), distance learning support systems (Anderson & Dron, 2011), intelligent learning support systems (VanLehn, 2011), and adaptive learning support systems (Khosravi et al., 2020). The integration of learning support systems with adaptive learning enhances the effectiveness of personalized learning, supporting learners across a range of subjects, including complex ones (Holstein et al., 2019).

Programming courses constitute a highly intricate subject within the fields of computer science and information technology, demanding considerable effort and problem-solving skills to overcome challenges effectively (Gomes & Mendes, 2007). Additionally, computer programming is a foundational subject essential for undergraduate students. Many novice learners frequently encounter challenges in understanding programming concepts, as solving programming problems necessitates both logical thinking and a foundational knowledge of programming (Cheung et al., 2010). For instance, some novices struggle to solve problems despite understanding the concepts due to a lack of knowledge on how to write code to address specific problems. Others face challenges when required to solve problems that necessitate the integration of multiple programming concepts, such as iteration, function, and data structure, to address a programming exercise. Although learners receive the same programming instruction, differences in their levels of understanding highlight the challenges some face in solving problems independently. These challenges are further compounded by limited classroom time and the need for guidance in selecting appropriate practice exercises.

To address these issues, this research aims to design, develop, and evaluate an adaptive learning support system that delivers exercises aligned with individual comprehension levels. To achieve this aim, we developed an ADaptiVe lEarning support system based on oNTology of mUltiple pRogramming languagEs (ADVENTURE) (Na Nongkhai et al., 2023), designed to provide programming exercises tailored to individual learners' skill levels. Moreover, this study aims to explore the integration of our previously developed ontology, CONTINUOUS (COmputatioNal ThinkIng oNtology mUltiple prOgramming langUageS) (Na Nongkhai et al., 2022), which encapsulates common concepts of multiple programming languages (including Python 3.x, Java, and C#), within ADVENTURE. Besides incorporating CONTINUOUS, the system integrates the Elo Rating System (ERS) in an educational setting (Pelánek, 2016) to support learners in practicing programming by recommending a suitable exercise. To explore the development of the proposed system and the effectiveness in learning outcomes, the following three research questions will be explored:

1. How can CONTINUOUS be utilized in programming support systems?
2. How can the Elo Rating System in an educational setting be adopted in an ontology-based adaptive learning system?
3. What disparities exist in learning performance—encompassing learning achievement and learning perception—between learners utilizing ADVENTURE's adaptive mode and those practicing with random exercise sequences?

To address the aim of designing and developing an ontology-based adaptive learning support system and evaluating its effectiveness, this study integrated CONTINUOUS with a real-time skill assessment mechanism based on the ERS in an educational setting. The system developed in this study, named ADVENTURE, supports learners by recommending programming exercises that are suitable for their skills. Its effectiveness is evaluated through an empirical study involving three experimental groups, focusing on learning performance. According to these components, the remainder of this paper is organized as follows: Section 2 reviews related work on programming ontologies and personalized, adaptive learning in computer programming. Section 3 presents the design and development of ADVENTURE, with a focus on integrating CONTINUOUS into the system. Section 4 details the incorporation of the Elo Rating System in an educational setting into the system's adaptive mechanisms. Section 5 describes the experimental methodology used in this study. Section 6 reports the results of the empirical study, including analyses of questionnaire responses and learning logs. Finally, Section 7 concludes the study by discussing key findings, limitations, and future work.

## 2. Related Work

### 2.1. An Ontology of Computer Programming

Guarino et al. (2009) define an ontology as a structured framework for a specific domain, encompassing a hierarchy of concepts, their interrelationships, and instances of each concept. Yun et al. (2009) extended the application of ontologies beyond linguistics to fields such as engineering and computer programming. They proposed a domain ontology for the C programming languages, featuring top-level classes such as "SYNTAX", "PROGRAMMING TECHNIQUE", and "PLATFORM", with concepts organized according to a course syllabus. Their findings demonstrated that the ontology clearly delineates the relationships among C programming concepts and holds promise for future reuse.

In addition to procedural concepts, several studies have utilized ontologies to represent object-oriented concepts within high-level programming languages. For instance, Kouneli et al. (2012) developed an ontology for Java that established relationships between fundamental and object-oriented concepts, serving as the foundation for a semantic web-based learning system at Hellenic Open University. Similarly, Protus 2.0 (Vesin et al., 2012) incorporated four ontologies—teaching strategy, learner model, task, and domain. The domain ontology, in particular, focused on Java programming concepts such as syntax, loops, and classes, providing the contextual knowledge within the Java tutoring system. Another example is OPAL (Cheung et al., 2010), an ontology-based framework for personalized adaptive learning in Java programming courses. OPAL dynamically adjusts the difficulty of subsequent topics based on learner performance, enabling learners to progress at an individualized pace. Shishehchi et al. (2021) also proposed a personalized learning system that used an ontology to recommend customized learning paths based on individual knowledge and goals. Their evaluation showed a significant reduction in the time required to find materials and highlighted the system's effectiveness in advancing personalized programming education.

In addition to the design of an ontology for a single programming language, Pierrakeas et al. (2012) proposed distinct ontologies for C and Java to facilitate online tutorials. Their work focused on foundational concepts and assisted instructors in visualizing the relationships between core concepts. Their analysis revealed numerous shared fundamental concepts, highlighting the potential for reusing instructional materials across these languages. Similarly, De Aguiar et al. (2019) introduced the OOC-O ontology as a unified framework for object-oriented programming, which was validated across multiple languages, including Java, C++, Python, Smalltalk, and Eiffel. This ontology emphasizes key principles such as abstraction and polymorphism. Khedr et al. (2021) developed a semantic query expansion system to improve the precision of search results for Java and Python by constructing specific ontologies for these languages, enhancing query disambiguation.

According to the studies referenced above, some research has developed ontologies of specific programming languages for computer programming education to enhance the understanding of the relations between programming concepts. However, research by Pierrakeas et al. (2012) and De Aguiar et al. (2019) has shown that most programming languages share similar concepts and syntaxes, as demonstrated through the design of ontologies supporting multiple programming languages. Additionally, the work on OPAL (Cheung et al., 2010) and the study by Shishehchi et al. (2021) have employed ontologies as knowledge graphs to facilitate adaptive and personalized learning paths. This demonstrates an approach to ontology's integration that goes beyond its traditional application in the semantic web. Consequently, we have adopted CONTINUOUS in ADVENTURE to support learners in practicing any programming language they need. The contribution of ontology's adoption in ADVENTURE will be discussed in Section 3.

## 2.2. Personalized and Adaptive Learning in Computer Programming

Personalized learning in computer programming pedagogy tailors learning processes and activities to meet individual learners' needs. Some studies have implemented personalized learning by recommending learning materials based on learners' style. For example, the Protus system (Klašnja-Milićević et al., 2011) provided Java programming materials tailored to individual learners by identifying their learning styles through an in-system survey. Although the Protus system followed the same sequence of programming topics for all learners, it personalized the learning materials to align with each learner's specific learning style, potentially enhancing their understanding of programming concepts.

Chookaew et al. (2014) expanded on this concept by developing a personalized e-learning environment for C programming. Their system generated independent sequences of programming concepts based on learners' problem areas and learning styles. For instance, the concept "variables and data types" could be followed by topics such as "input", "conditionals", "iteration", or "array structure", depending on the learners' needs. Their approach aimed to identify optimal topic sequences to facilitate learners' understanding. They used the percentage of incorrect answers for each concept to determine the appropriate sequence and gathered information about learning styles through a questionnaire. As a result, learners received both individualized learning materials and tailored concept sequences. Their finding demonstrated the effectiveness of this approach, showing a normalized gain of 0.70 between pre-and post-test scores.

Nevertheless, learners' understanding levels may fluctuate throughout the learning process, increasing or decreasing in response to different activities. Adaptive learning is one potential approach to address these varying levels of understanding in real time. Extensive research in computer programming education has explored the potential of adaptive learning approaches to enhance personalized learning experiences. For instance, Gavrilović et al. (2018) proposed an algorithm for adaptive learning in Java programming languages. In their system, each learning object included teaching materials, and task activities. The adaptive learning process was implemented in the task activities, where learners' code submissions were evaluated by a Java grader. After completing the learning materials, learners engaged in tasks with dynamically sequenced activities based on their individual performance. However, progression to the next learning object was contingent upon successfully completing the task activities within the current one. This approach enabled learners to study Java programming through instructor-developed materials and adaptive task-based activities. Additionally, the PyKinetic (Fabic et al., 2018), a mobile Python tutor, utilized adaptive problem selection across six Python topics and seven problem levels. The problems involved identifying errors, predicting output, and correcting code through multiple choice or drop-down list choices. The number of answer choices varied by the difficulty of the problems. The system applied an adaptive strategy to select the type and difficulty of questions based on individual learners' performance, which was measured by the time taken and the number of attempts. Furthermore, Troussas et al. (2021) proposed an adaptive teaching strategy for C# programming languages. Their system employed a normalized fuzzy weight to represent learners' knowledge levels, which informed decision-making for learning activities based on the Revised Bloom's Taxonomy (RBT). The decision-making rules were established by 15 faculty members who were experts in teaching computer programming. Each rule mapped to specific learning activities with varying degrees of complexity, depending on the learner's fuzzy weight score. The system evaluated learners' knowledge levels and delivered appropriate learning activities based on these adaptive rules.

Building on the previously discussed approaches, personalized and adaptive learning offers a diverse range of strategies to enhance computer programming education. These strategies include recommending appropriate learning materials, tailoring learning pathways based on individual needs, and dynamically adjusting activities in response to learner performance. However, most instructional materials in adaptive learning strategies focus on foundational theoretical concepts and problem-solving activities, such as coding exercises and multiple-choice programming problems. Frequent practice with a variety of programming problems is a well-established method for learners to develop and refine their programming skills (Amoako et al., 2013; Mbunge et al., 2021; Zhang et al., 2013). Consequently, ADVENTURE emphasizes practical programming by providing learners with programming problems tailored to their skill levels. A detailed discussion of these adaptive strategies will be presented in Section 4.

## 3. Adaptive Learning Support System Based on Ontology of Multiple Programming Languages

To address this first research question, we integrated CONTINUOUS within an adaptive learning support system, which has been developed in this study, called ADVENTURE. This integration is described in Sections 3.1, 3.2.3, and 3.4.

ADVENTURE is an adaptive learning support system for computer programming, developed as a web-based platform using React (https://reactjs.org/ accessed on 1 October 2022) and Spring Boot (https://spring.io/projects/spring-boot accessed on 1 October 2022). The advantage of using a web-based platform is that ADVENTURE enables learners to execute code without the need to install programming plugins, creating a conducive environment for practicing programming.

The procedural design of ADVENTURE prioritizes user-friendliness, as illustrated in Figure 1. There were five processes for utilizing ADVENTURE, detailed as follows:

The first step, as shown in ① in Figure 1. At their first use of the system, learners are required to complete a programming experience questionnaire. However, if this is not the first instance of use, learners are not required to complete this questionnaire.

The second step, as shown in ② in Figure 1. Based on their responses, the system presents a recommendation message suggesting an initial programming concept for learners with no prior programming experience, as shown in Figure 2. In contrast, learners with programming experience will not receive any recommendation, as shown in Figure 3, and can freely select any programming concept for practicing programming.

The third step, as shown in ③ in Figure 1. The system requires learners to complete pretests the first time they select each programming concept. This process helps determine the initial difficulty level for each concept. However, if learners select the same programming concept again, they can continue practicing at the recent difficulty level they achieved.

The fourth step, as shown in ④ in Figure 1. In this process, the system selects a programming question from the database by matching the learners' skills with exercise difficulty. The details of this matching and calculation are explained in Section 4.

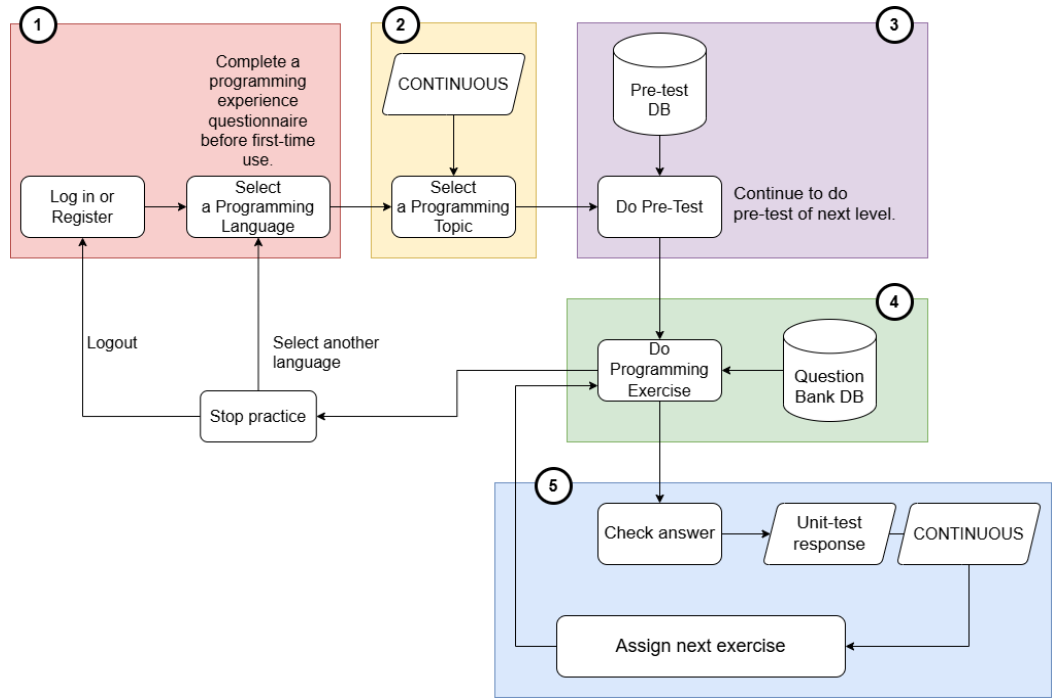

**Figure 1.** The process flow of ADVENTURE, including ① Select a programming language, ② Select a programming topic, ③ Do pre-test, ④ Do programming exercise, and ⑤ Receive the next programming exercise.

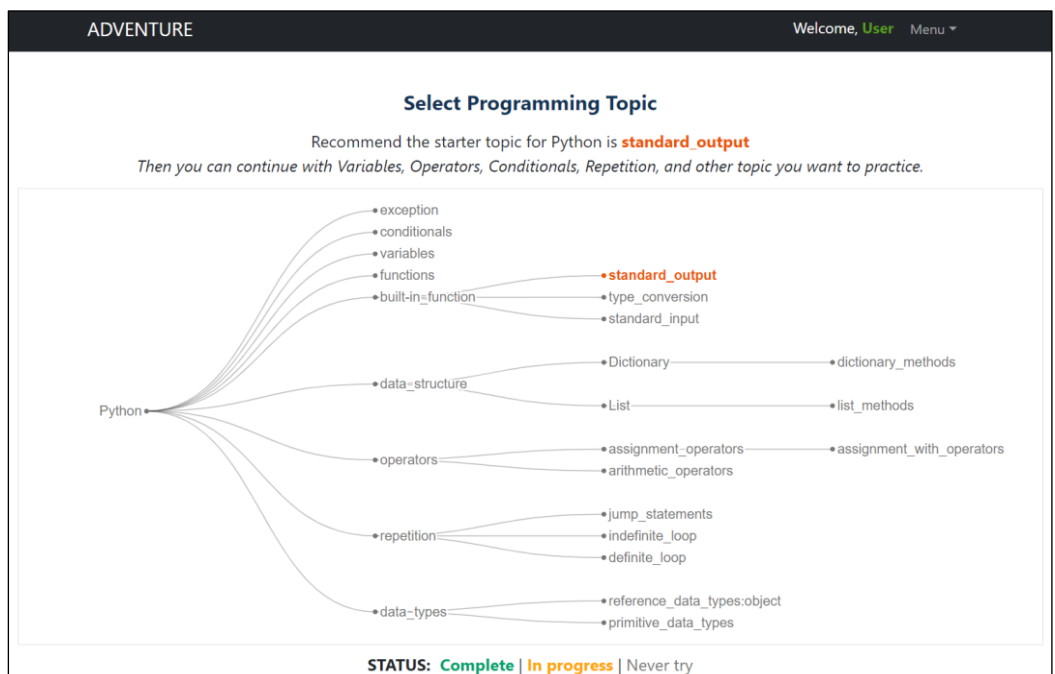

**Figure 2.** A suggested message of an initial programming concept for learners without prior programming experience.

The fifth step, as shown in ⑤ in Figure 1. This process includes an assessment of code submissions evaluated through a unit test. Additionally, it features a mechanism that recommends the next exercise, utilizing the ERS in an education setting to suggest future exercises.

### 3.1. Visualization of Programming Concepts

One of the ontology adoptions in (Na Nongkhai et al., 2023) is visualization programming concepts using CONTINUOUS (Figure 3). This development utilizes the tree layout of DTree (D3) (https://d3js.org/ accessed on 1 November 2022) to represent the hierarchy of programming concepts within a specific programming language. The system displays this hierarchy to help learners understand the relations between programming concepts and provides them with the flexibility to select any concept of interest. For instance, a learner may be interested in practicing the concept of "List". As shown in the programming concepts graph in Figure 3, the "List" concept encompasses a sub-concept known as "list_methods". In the event that a learner chooses to practice "list_methods" and encounters difficulties in completing the exercises, the system allows the learner to explore the related concept to "list_methods", to facilitate a necessary understanding.

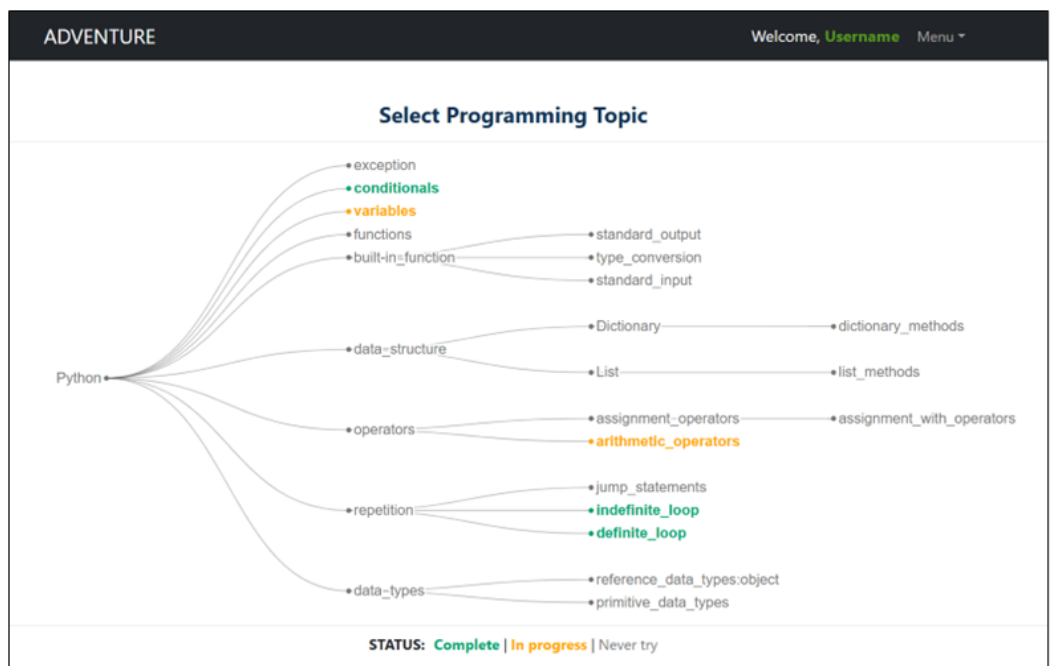

**Figure 3.** The visualization of programming concepts.

Additionally, learners can track their progress through color-coded statuses displayed at the bottom of the visualization. For example, a concept highlighted in orange indicates that it is currently in progress, as shown in Figure 3, where the topics "variables" and "arithmetic_operators" are shown to be in progress. A concept highlighted in green indicates that it is complete already, as shown in Figure 3, where the topics "conditionals", "indefinite_loop", and "definite_loop" are shown to be complete. This feature helps learners to easily monitor their progress during subsequent sessions with the system.

### 3.2. Programming Exercises

The programming exercise phase within ADVENTURE begins by presenting learners with a programming question aligned with their selected concept, allowing them to take the necessary time to complete the exercises. The webpage consists of four primary sections, as shown in Figure 4.

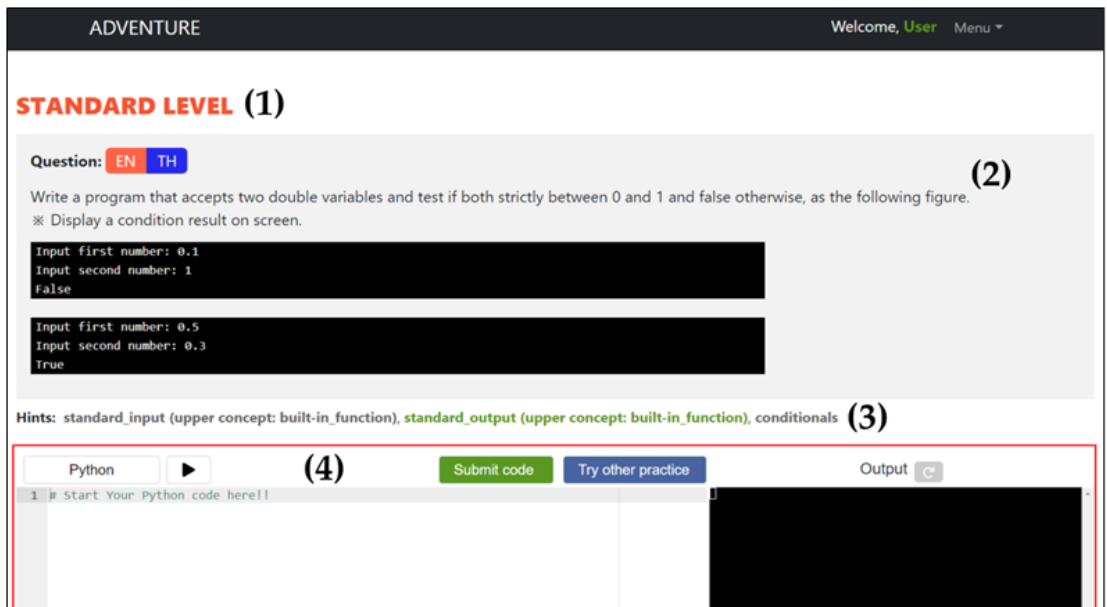

**Figure 4.** Interface of the programming exercise webpage, consisted of (1) Level of programming exercises, (2) Programming exercise Questions, (3) Hints for Programming Exercises, and (4) Programming Execution and Submission (within the red box).

3.2.1. Level of Programming Exercises

This research defines the difficulty levels of programming exercises as easy, standard, and difficult, as shown in (1) in Figure 4. The exercise level is determined through two methods. First, the initial level is assigned based on the results of a pre-test, which is provided by the system when a learner selects a programming concept for the first time. Second, the system adjusts the current level based on changes in the learners' skills in the current programming concept. When a learner achieves the required score (a threshold score of 0.85, which is adopted based on the study by Zheng et al. (2022); this detail has been described in Section 4.1), the system advances them to the next level. Conversely, if the learner's performance at the current level drops to zero, the system reassigns them to the previous level.

Moreover, when learners advance to the next difficulty level, their skills are reset to zero for that level. For instance, upon achieving or exceeding the threshold score in the easy level, learners will transition to the standard level, where their skills will once again begin at zero. In contrast, when learners are reassigned to a previous level, their most recent skill is retained. This allows learners to return to the current level after a single correct question, ensuring that temporary setbacks, such as an early incorrect answer, do not lead to long-term regression.

3.2.2. Programming Exercises Questions

This section contains programming questions and sample output curated by experienced instructors, accompanied by sample outputs for each question, as shown in (2) in Figure 4. The difficulty of each question is determined by the complexity of its code solution, which is one of the approaches utilized to assess exercise difficulty in Effenberger et al. (2019). When providing exercises, the system automatically assigns a question based on the alignment between the learners' skills and the difficulty of the exercises, which has been explained in Section 4.

### 3.2.3. Hints for Programming Exercises

Another application of ontology adoptions in (Na Nongkhai et al., 2023) involves providing hints for each programming question, as shown in (3) in Figure 4. These hints, defined by instructors based on the correct code answers and CONTINUOUS, highlight the selected concept in green and the related concepts in gray. Additionally, if a concept has a parent concept, it is automatically displayed within brackets; for example, the concept "standard_input" appears with its parent concept "built-in_function", as shown in (3) in Figure 4. This feature helps learners identify the concepts required to solve the question.

### 3.2.4. Programming Exercises and Submission

To evaluate their code, learners input their code into the editor (the white box on the left, as shown in (4) in Figure 4) and click the "Submit code" button to receive feedback. Additionally, learners can run their code by clicking on the "Run" button (located near programming language label), which displays the output in the terminal (the black box on the right, as shown in (4) in Figure 4). If learners encounter difficulty solving a problem, they can click the "Try other practice" button to receive other suitable questions. When this action is taken, the learner's skill will be marked down, similarly to the case when they submit incorrect answers. Their skill will be reduced based on the calculation of the ERS. Additionally, the skipped question will be recorded in the incomplete questions list. This list will appear after learners complete the selected concept. Furthermore, there is no limit on the number of skipped questions, as the system updates the learners' skills accordingly and provides the next exercise based on these updated skills.

## 3.3. Feedback from Code Submission

The assessment in ADVENTURE uses unit testing, which tests individual units of source code, and returns the results in JSON (JavaScript Object Notation) format. Upon the submission of learners' code, the system executes it using predefined test cases for each programming question, created by instructors. The system provides feedback for each test case, indicating whether the result is "Correct" or "Incorrect". For instance, consider a question that prompts learners to input two digits and check if both numbers are between 0 and 1 (Figure 5). This question consists of two test cases. If the learner submits the correct answer, the system indicates "Correct", as shown in Figure 6.

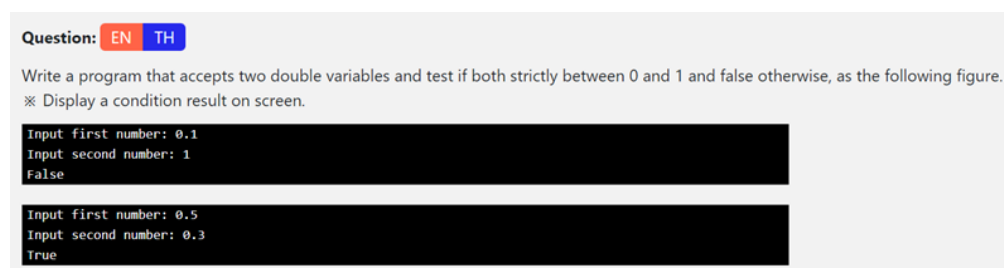

**Figure 5.** Example of a programming question which includes two languages: (1) EN for displaying a question in English, and (2) TH for displaying a question in Thai.

If the learner submits an incorrect answer, the system indicates "Incorrect", as shown in Figure 7. In such cases, the system displays the input data for each test case, the expected answer, and the learners' result. After reviewing the feedback, learners can proceed to the next exercise by clicking the "Click to the next exercise" button.

In another example involving multiple feedback responses within one assessment, consider a question that asks learners to input the age of a candidate and determine their eligibility to vote (Figure 8, assumed the output can be any number when a candidate is not eligible to vote). This question also includes two test cases. If the learner submits code that does not pass all the test cases, the system displays the "Correct" and "Incorrect" results, as shown in Figure 9.

**Figure 6.** Feedback for a "Correct" case.

**Figure 7.** Feedback for an "Incorrect" case, while the black box displays the correct output of a question, and the red box shows the output from a learner's code submission.

**Figure 8.** Example of a programming question which may cause multiple feedback responses within one assessment.

**Figure 9.** Feedback for "Correct" and "Incorrect" cases.

In the first test case, the input is "18", and the expected result is "True". However, the condition specified in the "if" statement on line 4 of the code submission (Figure 9) does not account for the value "18". As a result, the system marks the first test case as "Incorrect".

Moreover, all failed questions are compiled into a list of incomplete questions, as shown in Figure 10. This feature was designed to guide learners in identifying unresolved questions and allow them to attempt incomplete questions again.

**Figure 10.** Interface of a successful concept, displaying details for question number 58; this questions' detail depends on the question number when the cursor hovers over it.

### 3.4. Suggestion for Next Programming Concepts

After completing the difficult level of the selected concept, learners are directed to the successful concept webpage, as shown in Figure 10. At this stage, learners have two primary main options. First, those who prefer not to continue with the current concept can switch to a different concept by clicking the "Try another concept" button. Alternatively, learners who wish to continue practicing the current concept can either engage in a random exercise by clicking the "Do another random exercise" button or select a specific exercise by clicking on a question number. When a specific exercise is selected, the system displays its details when the cursor hovers over the question number, as shown in Figure 10. The question numbers are divided into two categories: questions that learners have not attempted and those they have attempted but have not yet succeeded in.

When learners click the "Try another concept" button, the system presents suggestions for the next concept to explore. These suggestions are drawn from the concepts referenced in each programming question, based on the CONTINUOUS ontology. The system generates a frequency list of concepts that are related to the current concept. From this list, the system identifies concepts that co-occur with the current concept (appearing together in one-to-one relationships). It then evaluates whether these concepts have not yet been attempted or are still in progress, and these are provided as concept suggestions.

For instance, if the current concept is "conditionals", the system retrieves all programming questions containing the "conditionals" concept, as shown in Figure 11. If other concepts frequently appear alongside "conditionals" in multiple questions, the system identifies these as the most related concepts. The individual concepts are collected, and their frequency of occurrences is counted, as shown in Table 1.

| editor_question_id | levels | topics |
|---|---|---|
| 18 | easy | conditionals |
| 19 | easy | conditionals,arithmetic_operators |
| 20 | easy | conditionals |
| 25 | easy | list,array,functions,repetition,conditionals,nested_control |
| 27 | easy | functions,definite_loop,conditionals,nested_control,standard_input |
| 33 | standard | exception,conditionals |
| 34 | standard | functions,conditionals,exception,nested_control |
| 35 | standard | functions,exception,jump_statements,conditionals,indefinite_loop |
| 37 | standard | repetition,conditionals,dictionary |
| 39 | standard | list,definite_loop,conditionals,jump_statements |
| 40 | standard | functions,repetition,conditionals,map,dictionary |
| 41 | standard | functions,conditionals |
| 42 | standard | array,conditionals,repetition |
| 43 | standard | repetition,conditionals,list_methods |
| 44 | standard | repetition,conditionals,list_methods,list,array |
| 45 | standard | repetition,conditionals,nested_control,dictionary_methods |
| 46 | standard | jump_statement,conditionals,assignment_with_operators,indefinite_loop |
| 48 | difficult | repetition,conditionals,dictionary,nested_control,map |
| 49 | difficult | jump_statements,conditionals,nested_control,arithmetic_operators,standard_input,definite_loop,list |
| 50 | difficult | conditionals,indefinite_loop,arithmetic_operators |
| 51 | difficult | indefinite_loop,conditionals,assignment_with_operators |
| 55 | difficult | functions,definite_loop,conditionals,assignment_with_operators |
| 56 | difficult | list_methods,jump_statements,conditionals,nested_control,repetition |
| 57 | difficult | definite_loop,conditionals,nested_control,dictionary,map |
| 58 | difficult | array_methods,list_methods,jump_statements,conditionals,nested_control,repetition |

**Figure 11.** List of programming questions containing the "conditionals" concept (The concepts that appeared in one-to-one relationship with the "conditionals" were highlighted in the red box).

Table 1 shows that 20 concepts are related to "conditionals", with 17 appearing in multiple questions. The system identifies which of those concepts appear one-to-one with "conditionals" (highlighted by the red box in Figure 11), including "arithmetic operators", "function", and "exception". From this selection, the system suggests concepts that the learner has not yet attempted or that are still in progress (Figure 12).

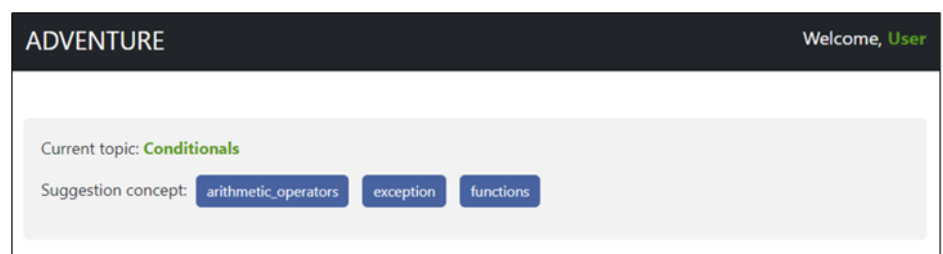

**Figure 12.** Example of a concept suggestion.

**Table 1.** Frequency of concept related to the "conditionals" concept.

| No. | Related Concept | Frequency |
|---|---|---|
| 1 | arithmetic_operators | 5 |
| 2 | array | 3 |
| 3 | array_methods | 4 |
| 4 | assignment_with_operators | 3 |
| 5 | definite_loop | 8 |
| 6 | dictionary | 4 |
| 7 | dictionary_methods | 1 |
| 8 | exception | 3 |
| 9 | functions | 11 |
| 10 | indefinite_loop | 2 |
| 11 | jump_statement | 1 |
| 12 | jump_statements | 5 |
| 13 | list | 3 |
| 14 | list_method | 1 |
| 15 | list_methods | 3 |
| 16 | map | 2 |
| 17 | nested_control | 14 |
| 18 | repetition | 13 |
| 19 | standard_input | 6 |
| 20 | standard_output | 2 |

In this example, the system prioritizes concepts that have a direct one-to-one relationship with "conditionals" and frequently co-occur with "conditionals" more than once, as in the suggestion concept that is shown in Figure 12. The concepts of "arithmetic_operators", "exception", and "functions" appear in a one-to-one relationship with "conditionals" (as shown in Figure 11). Additionally, the concepts of "arithmetic_operators", "exception", and "functions" also have frequencies of 5, 3, and 11, respectively (as shown in Table 1). However, if a concept that has a one-to-one relationship with the current concept appears only once, it will not be selected for the suggested concepts. This concept suggestion feature demonstrates another adoption of CONTINUOUS by guiding learners toward the next programming concepts based on their relevance and frequency.

## 4. The Elo Rating System in an Educational Setting Recommends Suitable Programming Exercises

To address the second research question in this study, we integrated the ERS in an educational setting within an adaptive learning support system. This section provides a detailed description of this integration process.

The Elo Rating System (ERS) is a statistical method for estimating a person's skills based on their performance history. Initially used in chess to facilitate pairing players based on skill levels (Elo, 1986), it has demonstrated its versatility by extending its utility across various sectors, including education. In educational settings, the ERS (Pelánek, 2016) is applied to estimate both learners' skills (the learners' current proficiency level that is dynamically updated based on their performance) and the item difficulties (the challenge of a question based on all learners' answers). For instance, Klinkenberg et al. (2011) illustrated the implementation of ERS in elementary mathematics education by providing students with math games that match their skill. To assign an appropriate programming exercise within ADVENTURE, we adopt a process based on ERS, as formulated by Pelánek (2016) for an educational setting, as follows:

$$\theta_s := \theta_s + K \cdot (correct_{si} - P(correct_{si} = 1)) \qquad (1)$$

$$d_i := d_i + K \cdot (P(correct_{si} = 1) - correct_{si}) \qquad (2)$$

In Formulations (1) and (2), we define $\theta_s$ as skills of a learner s, $d_i$ as the exercise difficulty of an exercise i; K as the learning rate (a weight value of learning) (Wauters et al., 2012); $correct_{si}$ as the correctness of a code submission for exercise i, where 1 represents a correct answer and 0 an incorrect one; and $P(correct_{si} = 1)$ as the probability of a correct answer, as shown in Formulation (3):

$$P(correct_{si} = 1) = \frac{1}{(1 + e^{-(\theta_s - d_i)})} \qquad (3)$$

The initial values for both learners' skills and exercise difficulties are set to zero. These values adjust based on learners' code submissions. A correct submission increases the learners' skill and decreases the exercise difficulty, whereas an incorrect submission lowers the skill and raises the difficulty of the exercise.

### 4.1. The Increment and Decrement of Learners' Skills and Exercise Difficulty

With each code submission, the system updates learners' skills and exercise difficulty levels, as illustrated in Figure 13, which shows the relationship between learners' skills and exercise difficulty upon submitting a correct answer. Initially, both learners' skills and exercise difficulty are set to zero. When learner-1 submits a correct answer to question-1, the system uses these initial values to calculate adjustments according to Formulations (1) and (2). This results in an increase in learner-1's skill level, indicated as "Q1" in the "Learners' Skill" graph, while the difficulty of question-1 decreases, shown as the first point in the "Exercise Difficulty" graph. When learner-2 submits a correct answer to question-1, the system then uses learner-2's initial skill level and the updated difficulty of question-1 to recalculate using the same equations. The difficulty of question-1 decreases further, shown as the second point in the "Exercise Difficulty" graph. Based on this example and the information shown in Figure 13, where the graphs represent scenarios of continuous successful completions, successfully answered questions will decrease in difficulty over time. Conversely, if learners are unable to complete questions, the difficulty level will increase.

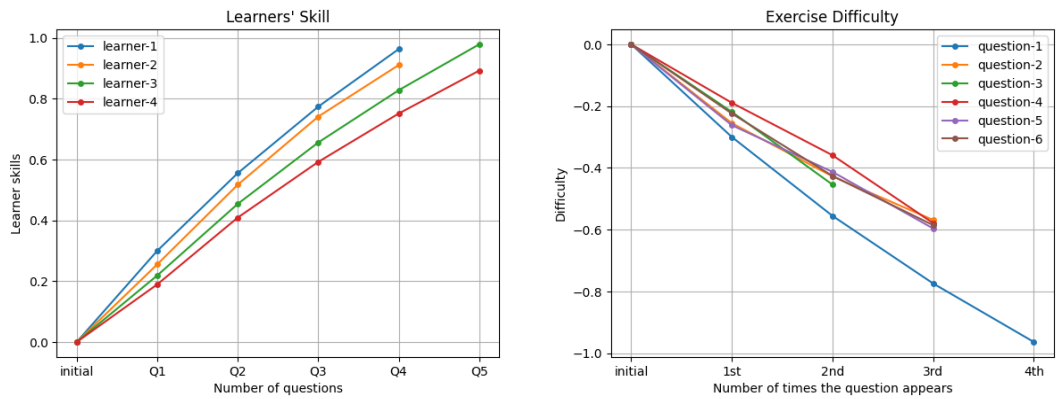

**Figure 13.** Relationship between (a) increment in learners' skills and (b) decrement in exercise difficulty.

Exercise difficulty is not restricted, while learners' skills range from 0 to 1. Learners' skill levels are evaluated across easy, standard, and difficult levels. When a learners' skill reaches or exceeds 0.85, which is the threshold score for advancement the system will promote the learner to the next level. This threshold score was adopted based on the study by Zheng et al. (2022), who proposed a knowledge mastery classification scheme in which scores equal to or above 0.85 indicated "Complete Grasp" of a concept. This classification informed the design of adaptive learning paths and exercise recommendations in their knowledge structure tree-based algorithm. Accordingly, we adopted the same threshold as the consistent criterion within our adaptive learning mechanism. When learners have reached the maximum level within the current programming concept, the system will direct them to the successful concept webpage (as detailed in Section 3.4 of this paper). Conversely, if the learner is at the minimum level, the system will continue to offer programming exercises at the easy level.

### 4.2. Matching Learners' Skills and Exercise Difficulties

To assign an appropriate programming exercise to learners, the system selects the question with the shortest distance between the learners' skills and the exercise difficulty. For instance, Figure 14 illustrates a scenario where five questions have varying difficulty levels, while the learners' skill level is assumed to start at approximately 0.36, reflecting a continuation of prior practice. When the learner re-engages with programming exercises, the system aims to identify the question with the shortest distance to the learners' current skill. Among the available questions, question-1 (denoted as (1) in Figure 14) is closest to the learners' skill, so the system assigns question-1. Upon successful completion of question-1, the learners' skill increases, the difficulty of question-1 decreases, and the system excludes question-1 from further consideration.

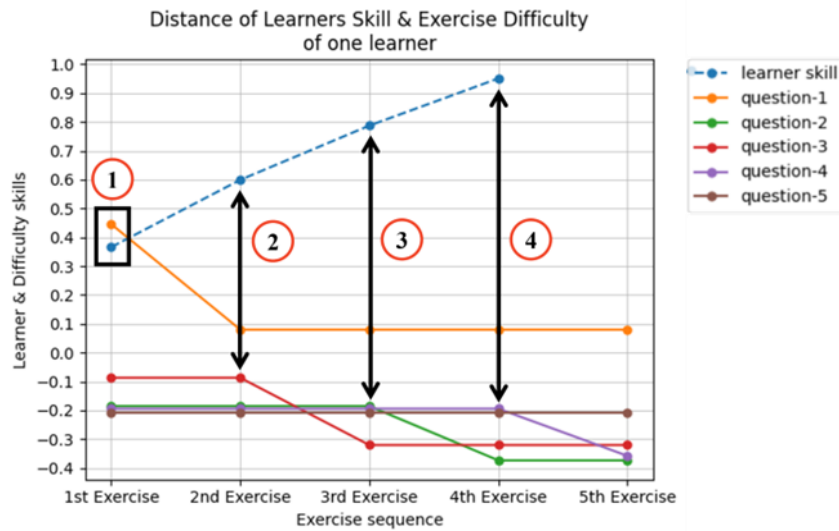

**Figure 14.** Matching learners' skill and exercise difficulty presents the following: ① The pairing of the current learners' skills and the question-1 difficulty, ② The pairing of the updated learners' skills and the question-3 difficulty, ③ The pairing of the updated learners' skills and the question-2 difficulty, and ④ The pairing of the updated learners' skills and the question-5 difficulty.

Following successful completion of question-1, the system recalculates and finds the shortest distance between the current learners' skill (approximately 0.6, as shown in (2) of Figure 14) and the remaining questions. As shown in (2) of Figure 14, question-3 is now the closest match, and the system assigns it to the learner. If the learner completes question-3 successfully, their skill increases, the difficulty of question-3 decreases, and it is excluded from future selection. This process continues for subsequent steps (3) and (4) in Figure 14.

After completing step (4), the learners' skill reaches approximately 0.95, meeting the successful criteria score for the level. If the current level is an easy or standard level, the system advances the learner to the next level, resetting their skill to the initial value for that level. However, at the difficult level, the system directs the learner to the successful concept webpage. As Figure 14 demonstrates, the learner does not complete question-5, as the required score is achieved within four questions. The number of exercises needed per level may vary depending on the learners' response in code submissions and the selected learning rate (K).

### 4.3. Selection of Learning Rate (K)

In the ERS developed in the 1960s, the parameter K represented the rating point value associated with a chess game score. Studies by Wauters et al. (2012) established K as the weight assigned to new observations, ensuring it is neither excessively large nor small. Pelánek (2016) identified the K as a key factor in the system's update rule. When K is set to a smaller value, the rate of change is gradual; a larger K results in faster adjustments. The value of K can be influenced by factors such as the number of responses or correct answers provided by learners, as well as the total number of responses within the system.

In this research, K is defined as the learning rate for adjusting learners' skills and exercise difficulty. The value is assigned based on the number of available questions at a given level. Figure 15 shows how the dynamic learning rate correlates with the number of correct answers. For example, with a learning rate of 0.1, learners need to complete over 11 questions successfully to reach the criteria score and advance to the next level. Conversely, a learning rate of 0.4 requires learners to complete six questions to progress, while a learning rate of 0.7 allows progression after just three correct answers. Requiring learners to complete numerous exercises to advance or complete concepts can lead to challenges that may impact engagement. However, progression is not based on the number of correct responses; it requires learners to achieve a threshold score derived from both the correctness and the expected success probability, as formulated in (1) and (3).

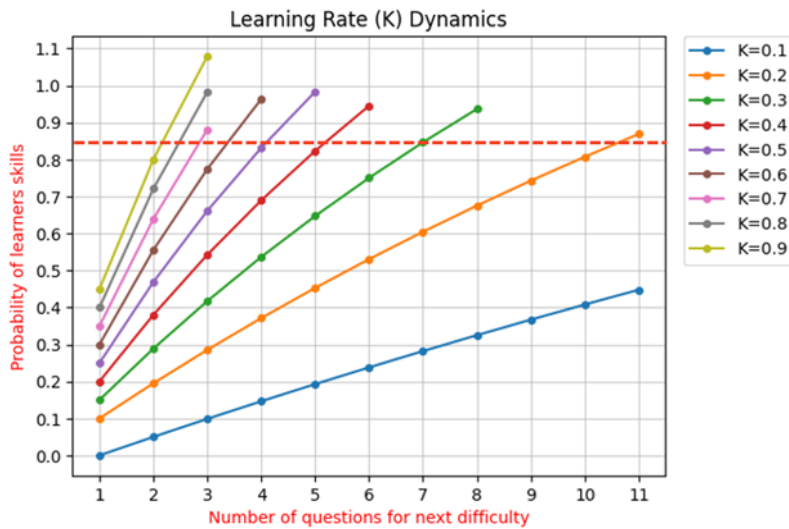

**Figure 15.** Influence of learning rate (K), while the threshold score is shown as a dash line.

Furthermore, incorrect submissions affect the value of the learning skills. As a result, repeated guessing without understanding does not facilitate advancement and might decrease the learning skill. This design discourages exhaustive trial-and-error strategies. Additionally, the number of questions per level in each concept varies: some levels include more than 10 questions, while others have fewer than 10. The selected learning rate ensures that learners are evaluated fairly across levels with differing content densities.

As shown in Figure 15, our analysis indicates that learning rates between 0.7 and 0.9 enable learners to progress to the next level after answering three questions correctly. Based on this finding, we conclude that a learning rate of 0.7 is suitable for situations where a minimum of three questions must be answered correctly. This threshold represents the minimum requirement for progression. However, if the programming concept at the current level consists of only three questions, this would require learners to achieve a 100% success rate, which could present a substantial challenge for novice learners in their programming practice. Therefore, it is recommended that each level within every concept comprises a minimum of four or five questions. This setup allows learners to succeed in approximately 60% to 75% of the questions, aligning with the minimum success rate for progression.

Further analysis reveals that a learning rate of 0.3 requires learners to answer eight questions correctly, while a learning rate of 0.4 allows advancement after six correct answers. Based on this finding, we conclude that a learning rate of 0.4 is appropriate for scenarios where the objective is to answer a maximum of six questions correctly. This threshold represents the maximum requirement for progression. For a learning rate of 0.5, learners need to answer five questions correctly. The learning rate gradually increases until it aligns with the threshold number of correct answers for a rate of 0.7.

According to the definitions provided of the minimum and maximum learning rates, we have drawn conclusions regarding progression, which are represented in Table 2. Therefore, the criterion for selecting the appropriate learning rate is based on the number of questions at each difficulty level of the selection concept. For example, consider the easy level of the

"variables" concept, which includes eight questions. At this level, learners are required to answer five questions correctly, with a learning rate of 0.5. This learning rate will be applied in the K parameters of Formulations (1) and (2) to update both learners' skills and the exercise difficulty. If a level comprises more than nine questions, the system will apply a learning rate of 0.4 to adjust learners' skills and the exercise difficulty.

**Table 2.** Selection of learning rate ($K$).

| Learning Rate (K) | Number of Pass Question (N) | Total of Question Number |
|---|---|---|
| 0.7 | 3 | 4–5 |
| 0.6 | 4 | 6 |
| 0.5 | 5 | 7–8 |
| 0.4 | 6 | More than or equal 9 |

## 5. Experimental Methods

### 5.1. Measurements Techniques

The measurement technique employed in this study includes learning achievements from learning logs of the system (e.g., pre-test, code submission) and learner response to a questionnaire designed to evaluate learning-related perceptions. The tests were developed by programming instructors with over ten years of experience in computer programming education, and the questionnaire was developed based on the study by Wang et al. (2014, 2020), with guidance from experienced instructors. The questionnaire contains 25 questions rated on a seven-point Likert scale (1–3: strongly to slightly disagree, 4: neutral, and 5–7: slightly to strongly agree). The questions are grouped into four main categories: mental effort (2 questions, $\alpha = 0.542$), encompassing efforts for understanding of experiment purpose and learning activities; mental load (3 questions, $\alpha = 0.669$), encompassing distraction and pressure, with distraction involving the selection of programming concepts and execution of exercises; technology acceptance (10 questions, $\alpha = 0.867$), involving five features in the system in terms of ease of use and usefulness; and satisfaction (10 questions, $\alpha = 0.912$). The questionnaire will be administered to participants at the end of the experimental period.

### 5.2. Participants and Experimental Procedure

The participants in this research were 56 undergraduate students from the Information Technology faculty of a university in Thailand. The participants were divided into three groups, each consisting of individuals aged 20–22: a preliminary group, an experimental group, and a control group.

The preliminary group consisted of 15 third-year undergraduate students (12 males and 3 females) who had completed seven programming sessions. Each session included 45 h of lectures and 45 h of practical activities prior to the experiment.

Subsequently, the experimental and control groups involved 41 second-year undergraduate students. These participants were divided into two groups based on the natural class setting of a computer programming course taught by the same instructor. Both groups had completed four programming sessions. Each session included 45 h of lectures and 45 h of practical activities prior to the experiment. One class, comprised of 26 students (14 males and 12 females), was assigned as the experimental group, while another class consisting of 15 students (15 males) was assigned as the control group. Since the groups were assigned by natural class setting, controlling the gender distribution was challenging, affecting the

unbalanced composition of the control group. (This limitation will be discussed at the end of this chapter.)

Figure 16 shows the experimental procedure, which consisted of two phases. In the first phase, the preliminary group engaged with ADVENTURE 1.0 for one week, both during and outside the practical programming class. They spent one hour engaging in programming exercises using the adaptive mode of ADVENTURE 1.0. Additionally, they were allowed to continue practicing programming exercises outside of the classroom for the following week. Upon completion of this practice period, they completed a questionnaire designed to capture their learning-related perceptions and provide feedback on the system. Based on this feedback, ADVENTURE 1.0 was updated to version 1.1.

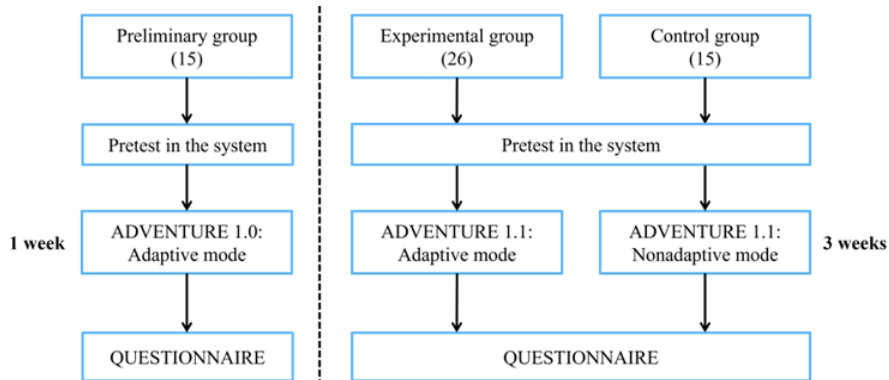

Figure 16. Experimental procedure.

In the second phase, the experimental and control groups engaged ADVENTURE 1.1 for three weeks, utilizing the system both during programming class and through independent study outside the classroom. This phase involved the use of distinct versions of ADVENTURE 1.1. The experimental group used ADVENTURE 1.1 in adaptive mode, which delivered programming exercises based on updates of learners' skills and exercise difficulty calculated by the ERS in an education setting. In contrast, the control group used ADVENTURE 1.1 in non-adaptive mode, which delivered exercises by randomly selecting all the questions in the programming concept selection without any adaptive reasoning. The participants in the control group received random exercises of mixed difficulty levels and were not informed about the difficulty level of each question.

For example, when a participant from the experimental and control groups selected the "definitie_loop" concept, the participant in the experimental group received an exercise at an order of difficulty level that was assessed in the adaptive mechanism. They could continue and complete this concept with the threshold score outlined in Section 4.1. However, the participant in the control group was randomly assigned exercises that varied in difficulty. To achieve success in each concept, they were required to provide correct answers for more than 60% of all the questions in that concept (this percentage refers to the setup percentage for defining the learning rate outlined in Section 4.3).

Finally, at the end of the experiment, participants in the experimental and control groups were required to complete the questionnaire to provide feedback on their experience using the system.

During the experiment period, participants' learning behaviors were recorded in learning logs. These logs consisted of a programming experience questionnaire completed before using any system functions, participants' pre-test scores of each programming concept, progress in practicing programming exercises, their skill levels for each concept and difficulty level, and details of their code submissions. The code submissions data included the learners' code, the execution result, and the time taken to complete each exercise. We collected a total of 1186 code submissions from all three groups: the preliminary group (127 submissions), the experimental group (936 submissions), and the control group (123 submissions). These datasets were utilized to analyze the learners' performance in programming skills.

There were some limitations in this experiment. Firstly, although the preliminary group comprised third-year students with more extensive programming experience than the second-year students in the experimental and control groups, their learning outcomes were compared in the overall analysis. The preliminary group primarily served to provide formative feedback on the initial version of the system (ADVENTURE 1.0), which was instrumental in guiding the development of ADVENTURE 1.1. Their inclusion also enabled an exploratory examination of the system's usability and potential effectiveness across varying levels of programming experience. In contrast, the experimental and control groups were composed of students with comparable programming experience and were taught by the same instructor, ensuring equivalent instructional conditions. Accordingly, the analysis comparing the adaptive and non-adaptive modes focused exclusively on the experimental and control groups, while the preliminary group was analyzed in relation to system usability and cross-level applicability.

Secondly, although the study design controlled for key variables, such as academic level, instructor, learning content, and exercise set, the adaptive mode was the only factor that differed between the experimental and control groups. This design aims to isolate the effects of the adaptive mode on learner performance. However, the small sample size may limit the generalizability of the findings. In addition to the small sample size, the last limitation is an imbalance among the experimental groups. The participants were selected from natural programming classes. Two classes, taught by the same instructor, were assigned as experimental and control groups in the second phase, respectively. Consequently, this selection method made it challenging to achieve an equal number of participants and a balance of genders in each group.

## 6. Results

To address the third research question in this study, we analyzed learners' perceptions from the questionnaire and learners' performance from learning logs.

### 6.1. The Analysis of Questionnaire Responses

Table 3 shows a summary of the analysis results based on the questionnaire response from 56 participants across all groups.

**Table 3.** Feedback of learning perceptions in the questionnaire.

|  |  | Mental Effort | | Mental Load | | Technology Acceptance | | Satisfaction |
|---|---|---|---|---|---|---|---|---|
|  |  | **Purpose** | **Learning Activity** | **Distraction** | **Pressure** | **Ease of Use** | **Usefulness** |  |
| Preliminary group | Mean | 3.80 | 4.73 | 3.60 | 3.73 | 5.53 | 5.59 | 5.28 |
| (N = 15) | S.D. | 1.52 | 1.53 | 1.06 | 1.71 | 1.02 | 0.99 | 1.21 |
| Experimental group | Mean | 3.46 | 5.00 | 3.56 | 3.50 | 5.58 | 5.85 | 5.76 |
| (N = 26) | S.D. | 1.50 | 1.50 | 1.28 | 1.17 | 0.74 | 0.65 | 0.73 |
| Control group | Mean | 4.47 | 5.27 | 4.17 | 4.07 | 5.48 | 5.63 | 5.51 |
| (N = 15) | S.D. | 1.36 | 1.62 | 1.18 | 1.44 | 1.19 | 0.87 | 1.01 |
| MANOVA | F | 1.229 | | 0.629 | | 1.306 | |  |
| (Wilks' Lambda) | P | 0.303 | | 0.707 | | 0.197 | |  |
| One-way ANOVA | F | 2.224 | 0.449 | 1.353 | 0.777 | 0.059 | 0.654 | 1.226 |
|  | P | 0.118 | 0.641 | 0.267 | 0.465 | 0.943 | 0.524 | 0.302 |

Note: df(2,53) in MANOVA.

For "mental effort", the average effort rating to comprehend the purpose of utilizing the system in the preliminary and experimental groups is 3.80 (S.D. = 1.52) and 3.46 (S.D. = 1.50), respectively, which is below the neutral point of 4. In contrast, the average rating for the control group reaches the neutral point of 4.47 (S.D. = 1.36). This indicates that participants who utilized the adaptive mode in the system easily grasped the purpose of system usage with moderate mental effort compared to those who engaged in random exercises. Meanwhile, the average effort ratings across all the experimental groups, which slightly exceeds the neutral point, indicate that the exercises presented a slight difficulty level for participants. However, MANOVA and ANOVA results suggest there were no significant differences between groups.

With regard to "mental load", the average rating of system distraction in both the preliminary group (Mean = 3.60, S.D. = 1.06) and experimental group (Mean = 3.56, S.D. = 1.28) are below the neutral point, and the control group is slightly higher than the neutral point (Mean = 4.17, S.D. = 1.18). This suggests that neither the visualization of programming concepts nor the exercise description on the webpage distracted learners from selecting concepts or engaging in practice. Similarly, the average pressure rating for utilizing the system in both the preliminary group (Mean = 3.73, S.D.= 1.71) and experimental group (Mean = 3.50, S.D. = 1.17) are below the neutral point, while the control group reaches the neutral point of 4.07 (S.D. = 1.44). This indicates that participants who practiced random exercises were slightly more challenged than participants who practiced exercises with the adaptive mode. However, MANOVA and ANOVA results suggest there were no significant differences between all the groups.

Regarding "technology acceptance", the average rating on the ease of use of the five system features—including the visualization of programming concepts, the overall webpage of the programming exercise, the hints on the exercise webpage, the feedback from code submission, and the suggestions for the next programming concept—is 5.53 (S.D. = 1.02) for the preliminary group, 5.58 (S.D. = 0.74) for the experimental group, and 5.48 (S.D. = 1.19) for the control group. Similarly, the average rating of the usefulness of these features is 5.59 (S.D. = 0.99) for the preliminary group, 5.85 (S.D. = 0.65) for the experimental group, and 5.63 (S.D. = 0.87) for the control group. All experimental groups' average ratings on ease of use and usefulness exceeded the neutral point. This indicates that the system is user-friendly and beneficial for all participants. Moreover, MANOVA and ANOVA results suggest there are no significant differences between all the groups in the five system features in terms of technology acceptance.

For more details of "technology acceptance" in each system feature:

First, the graph visualization of programming concept selection represents the first contribution of CONTINUOUS's integration into ADVENTURE. The average "ease of use" rating from all participants was 5.89 (S.D. = 0.95), and the "usefulness" rating was 5.91 (S.D. = 0.82). These results suggest that many participants slightly agreed or agreed (with an average rating of nearly 6) regarding the ease of use and the facilitation of programming concept selection through graph visualization. These quantitative results are supported by quantitative feedback from participants, as presented in Table 4. For instance, comments such as P-SP1 and P-SE1 from the preliminary and experimental groups highlighted the clarity of the concept graph's structure. Additionally, P-SE2 and P-SC1, from the experimental and control groups, reflect participants' appreciation for the ease of navigation. These consistent quantitative ratings and feedback across all groups suggest that this feature contributes to learners' perceived ease of use and conceptual clarity, regardless of whether they participated in adaptive or random practice. This suggests that this feature did not have any perceptual disparities between the adaptive and random practice conditions.

Second, for the overall webpage of the programming exercise, the average "ease of use" rating was 5.57 (S.D. = 1.19), and the "usefulness" rating was 5.59 (S.D. = 1.17). These results indicate that many participants slightly agreed that the entire webpage was easy to understand and useful. The qualitative feedback further supports these findings, as shown in Table 4. Participants from all groups highlighted the system's user-friendly interface (P-SP2) and its support for multiple programming languages (P-SE3 and P-SC2). However, some usability issues were noted, particularly concerning the Execute Code button (the white button with a triangle symbol, as shown in (4) in Figure 4). Some participants in the preliminary and experimental groups reported having difficulty locating the run code button, citing its lack of visibility. They suggested using icons and color to enhance the clarity of the interface (N-SP1 and N-SE1). In contrast, the control group reported no issues with the user interface. The findings suggest that while some participants in the adaptive mode experienced minor user interface issues, the consistency in qualitative ratings and feedback demonstrates that this feature facilitated participants in practicing multiple programming languages, regardless of whether they engaged in adaptive or random practice. Consequently, it appears that this feature had a minimal perceptual disparity between the adaptive and random practice conditions.

Third, for the hints that were provided within each programming question, the average "ease of use" rating was 5.09 (S.D. = 1.67), and the "usefulness" rating was 5.54 (S.D. = 1.24). These findings reveal that many participants slightly agreed that the hints help them identify essential programming concepts for problem-solving. Qualitative feedback supports these results, as shown in Table 4. A participant in the experimental group (P-SE4) indicated that the hint feature positively influenced their learning experience. This suggests that the hints provided targeted scaffolding, which allowed learners to focus on essential concepts without being encumbered by uncertainty. However, some participants from both the experimental and control groups expressed a desire for more comprehensive hint options (N-SE2 and N-SC1). Although the qualitative ratings and feedback indicate that the existing hint system was generally perceived as beneficial, participants from both the adaptive and random practice groups requested an increase in the quantity of hints with the learners' contextual needs. This consistent perspective across both adaptive and random practice groups suggests that this feature did not have any perceptual disparities between them.

Table 4. Qualitative comments from participants in the questionnaire.

| Group | Positive | Negative |
|---|---|---|
| Preliminary group | P-SP1: "A key benefit is the ability to clearly visualize the overall structure of programming topics".<br>P-SP2: "User-friendly interface with support for multiple programming languages".<br>P-SP3: "Highlighting errors along with explanations makes it easier for beginners to use". | N-SP1: "The run button should be more noticeable, and buttons should have icons and meaningful colors rather than just text". |
| Experimental group | P-SE1: "The concept graph is easy to understand".<br>P-SE2: "The sequence is easy to follow".<br>P-SE3: "Supports multiple programming languages and provides a free space to write code without needing additional installation". | N-SE1: "I couldn't find the run code button".<br>N-SE2: "More hints, please".<br>N-SE3: "Improve test cases to allow for more flexible answer formats". |

| Group | Positive | Negative |
| --- | --- | --- |
| | P-SE4: "Hints help guide the programming process, reducing the cognitive load of figuring out what to use". <br> P-SE5: "This helped me understand the concepts better". <br> P-SE6: "This suggesting concept function is easy to understand and very useful". | |
| Control group | P-SC1: "I like that the content is categorized clearly into sections". <br> P-SC2: "The code testing interface works regardless of programming language". <br> P-SC3: "No issues in the suggesting concept function, because it's beneficial". | N-SC1: "I would like a few more hint features to be added". <br> N-SC2: "Partial solutions should be provided to make the problem easier to understand". <br> N-SC3: "The level structure should be clearer". <br> N-SC4: "The system should be divided into levels". <br> N-SC5: "It would be helpful to have learning levels, such as Beginner, Experienced, and Review Mode". |

Note: Abbreviations are used to indicate the type of feedback and the participant group. P = Positive feedback, and N = Negative feedback. SP = Students in the Preliminary group, SE = Students in the Experimental group, and SC = Students in the Control group. For example, P-SC refers to positive feedback from students in the Control group, and N-SE refers to negative feedback from students in the Experimental group.

Fourth, for the feedback from code submissions, the average "ease of use" rating was 5.34 (S.D. = 1.43), and the "usefulness" rating was 5.82 (S.D. = 0.96). These results suggest that many participants slightly agreed that the feedback was easy to understand and either slightly agreed or agreed that this system's feature was useful. These quantitative findings are supported by quantitative feedback, as presented in Table 4. Participants from the preliminary group emphasized the clarity and usefulness of the system's error explanations (P-SP3). However, some participants from the experimental and control groups provided constructive suggestions for improvement, including the explanation of partial solutions to facilitate better understanding (N-SC2) and implementing a more flexible test case handling to accommodate multiple correct solutions (N-SE3). The qualitative ratings and feedback indicate that the current feedback system effectively provides clear explanations for beginners. However, feedback from participants engaged in both adaptive and random practice suggests that the system could be further enhanced by incorporating more detailed and adaptive feedback mechanisms. This consistent perspective suggests that this feature did not have any perceptual disparities between the adaptive and random practice conditions.

Fifth, for the suggestions for the next programming concepts, the average "ease of use" rating was 5.82 (S.D. = 1.08), and the "usefulness" rating was 5.75 (S.D. = 1.07). These results suggest that many participants slightly agreed or agreed that the suggestions helped them understand the relationships between programming concepts. Qualitative comments further support these findings, as presented in Table 4, with participants from the experimental group emphasizing improved concept understanding and ease of use (P-SE5 and P-SE6), while a control group participant reported the benefit of this feature (P-SC3). These results suggest that this feature effectively assisted participants in navigating to the next programming concepts. Additionally, feedback from participants who engaged in both adaptive and random practice reported that this feature was easy to understand and useful. This suggests that this feature did not have any perceptual disparities between the adaptive and random practice conditions.

Regarding "satisfaction", the average rating of general satisfaction with the system is 5.28 (S.D. = 1.21), 5.76 (S.D. = 0.73), and 5.51 (S.D. = 1.01) for the preliminary, experimental, and control groups, respectively. This suggests that most participants in each group were satisfied with engaging in programming practice within the system. Although the average rating across all groups was not very different, there was some negative feedback from some participants in the control group regarding exercise difficulty, suggesting the needs for a clearer level structure and the inclusion of defined learning levels such as Beginner, Experienced, and the need for the inclusion of a Review Mode based on the qualitative feedback from N-SC3, N-SC4, and NSC5, as presented in Table 4. These participants clearly experienced slightly more pressure due to the randomly assigned exercises in non-adaptive mode. However, the ANOVA result suggests there were no significant differences between all the groups. Nevertheless, the MANOVA was not reported in terms of "satisfaction", as it was measured using 10 distinct questions, which makes MANOVA not suitable to be used. Therefore, univariate ANOVAs were conducted for each item to enable a more detailed analysis of individual satisfaction dimensions and no significance is discovered.

### 6.2. The Analysis of Learning Logs in ADVENTURE

The learning logs from the three experimental groups contain two primary categories of features that highlight significant differences between them, as illustrated in Figures 17 and 18. First, the characteristics of code submissions were categorized into four primary features, as shown in Figure 17: (1) The submission of correct answers. (2) The submission of incorrect answers. (3) The submission of code without the necessary programming logic. (4) The frequency of requesting another programming exercise. Second, they were characterized by the rate of successful and unsuccessful programming concepts, as shown in Figure 18.

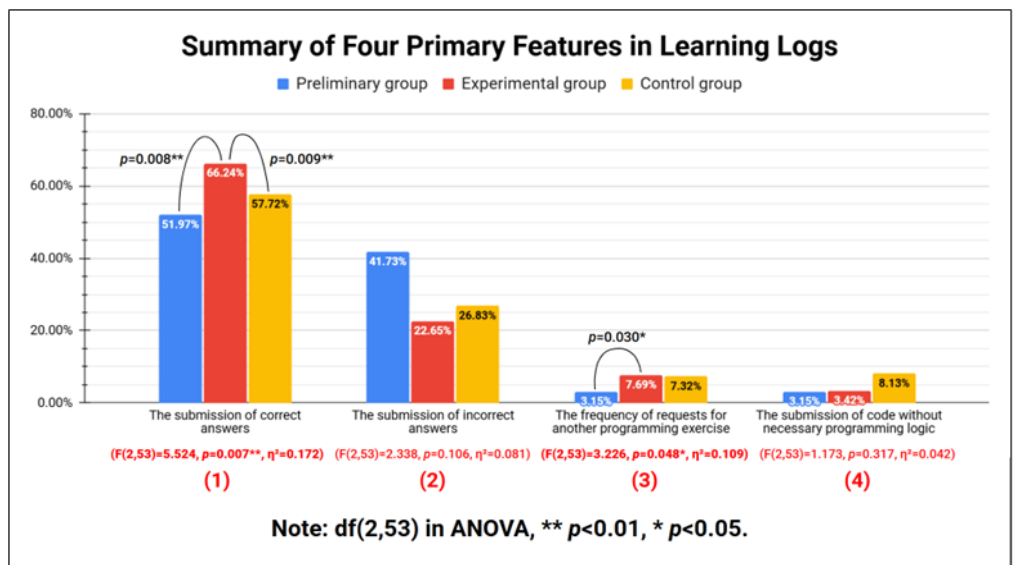

**Figure 17.** The bar chart titled "Summary of Four Primary Features in Learning Logs" from ADVENTURE presents the following: (1) The submission of correct answers ($F(2,53)=5.524$, $p=0.007**$, $\eta2=0.172$). (2) The submission of incorrect answers ($F(2,53)=2.338$, $p=0.106$, $\eta2=0.081$). (3) The frequency of requests for another programming exercise ($F(2,53)=3.226$, $p=0.048*$, $\eta2=0.109$). (4) The submission of code without the necessary programming logic ($F(2,53)=1.173$, $p=0.317$, $\eta2=0.042$).

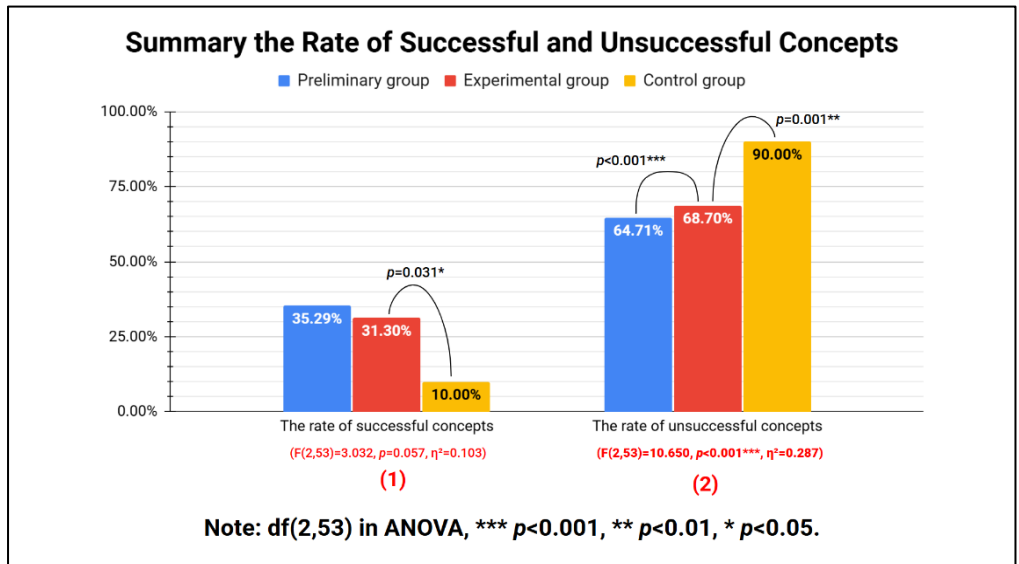

**Figure 18.** The bar chart titled "Summary the Frequency of Successful and Unsuccessful Concepts" from ADVENTURE presents the following: (1) The rate of successful concepts (F(2,53)=3.032, p=0.057, η2=0.103). (2) The rate of unsuccessful concepts (F(2,53)=10.650, p<0.001***, η2=0.287).

Among all the features presented in the bar charts, three demonstrated significant differences between the groups: First, the submission of correct answers (F(2,53)=5.524, p<0.01, η2=0.172), as shown in (1) in Figure 17. Second, the frequency of requesting another programming exercise (F(2,53)=3.226, p<0.05, η2=0.109), as shown in (3) in Figure 17. And finally, the rate of unsuccessful concepts (F(2,53)=10.650, p<0.001, η2=0.287), as shown in (2) in Figure 18.

For "the submission of correct answers", the experimental group demonstrated a significant higher percentage of correct submissions compared to both the preliminary group and the control group, as indicated by a significant overall ANOVA result (F(2,53)=5.524, p<0.01, η2=0.172). The large effect size suggests that the group distribution had a substantial impact on performance. Further pairwise comparisons showed that the experimental group significantly outperformed the preliminary group (p=0.008, ΔM =19.45). This finding indicates that although the preliminary group had greater prior programming experience, the updated version of ADVENTURE delivered skill-based programming exercises more effectively than the initial version. Additionally, the experimental group significantly outperformed the control group (p=0.009, ΔM =19.11) indicating that participants who received the random difficulty exercises may have found it more challenging to submit correct answers than those who received exercises adapted to their skill levels.

"The frequency of requesting another programming exercise" refers to instances where learners want to abandon the current programming question and attempt an alternative one instead. As shown in (3) in Figure 17, the experimental group requested another exercise significantly more frequently than the preliminary group based on a significant overall ANOVA result (F(2,53)=3.226, p<0.05, η2=0.109). Although the overall ANOVA results showed moderate effect sizes, the pairwise comparison between the experimental and preliminary groups was not statistically significant (p=0.03, ΔM =2.50). This suggests that the preliminary group, which had more prior programming experience, was less likely to abandon the current question compared to the experimental group, which had a similar level of programming experience to the control group. According to the overall and pairwise ANOVA results, which had moderate and small effect sizes, respectively, prior programming experience appears to affect the likelihood of requesting another programming exercise more than the learning approach.

For "the rate of unsuccessful concepts", the experimental
group had significantly higher unsuccessful concepts compared to the preliminary group ($p<0.001$, $\Delta M =2.73$), as indicated by a significant overall ANOVA result ($F(2,53)=10.650$, $p<0.01$, $\eta2=0.287$) possibly because participants in the preliminary group had more prior programming experience. Additionally, "the rate of successful concepts" shows that there were no significant differences between the preliminary and experimental groups, based on the overall ANOVA result ($F(2,53)=3.032$, $p=0.057$, $\eta2=0.103$). This indicates that prior programming experience affects the success rate with a moderate effect size.

In contrast, another pairwise comparison demonstrates that the experimental group had a significantly lower rate of unsuccessful concept completion than the control group ($p=0.001$, $\Delta M =-2.26$). Moreover, regarding "the rate of successful concepts", the experimental group demonstrated a significantly higher success than the control group ($p=0.031$, $\Delta M =1.44$) based on the potentiality significance of OVA result ($F(2,53)=3.032$, $p=0.057$, $\eta2=0.103$). This indicates that participants who received random exercises were more likely to abandon a challenging concept and attempt another one instead, whereas those who received adaptive exercises were more inclined to complete the programming concepts.

In summary, the results presented in Figure 17 show that the experimental group had significantly more correct submissions than the control group. This finding suggests the effectiveness of the adaptive mechanism in supporting programming practice. Furthermore, although differences in submissions lacking programming logic were not statistically significant, the experimental group submitted such code less frequently than the control group. This suggests that the random exercise assignments in the control group could have led to skill–task misalignment, prompting a reliance on guessing and resulting in less logically structured code, resulting in a higher frequency of submissions lacking the necessary logical structure compared to the experimental group.

Regarding the results presented in Figure 18, in addition to the significant lower rate of unsuccessful concepts comparing to control group, it is found that participants in the experimental group demonstrated significantly greater success in completing programming concepts compared to those in the control group. This finding suggests that the adaptive learning mode may enhance learner motivation in completing the programming concepts through exercises that align with their individual skill levels. In contrast, participants in the control group, who practiced with randomly assigned exercise difficulties, may have experienced greater effort in practicing programming (as shown in table 3), potentially leading to abandoning the completion of programming concepts.

## 7. Discussion and Conclusions

### 7.1. Discussion and Implications

This research proposes the development of ADVENTURE to support learners in improving their programming skills through the integration of CONTINUOUS and the ERS in an educational setting. Several studies have developed an ontology for a single programming language to clarify conceptual relationships (Yun et al., 2009; Kouneli et al., 2012), while others have applied ontologies in tutoring systems and personalized learning (Vesin et al., 2012; Cheung et al., 2010). Some studies have expanded this by modelling relationships across multiple languages (Pierrakeas et al., 2012; De Aguiar et al., 2019; Khedr et al., 2021), demonstrating that ontology-based knowledge graphs effectively represent interrelated programming concepts. Building on our prior work (Na Nongkhai et al., 2022)—which proposed the development of CONTINUOUS—and the advantages of the above studies, this study successfully adopted CONTINUOUS into ADVENTURE. This integration led to three contributions that addressed the first research question: "How can CONTINUOUS be utilized in a programming support system?" The first contribution is the visualization of programming concepts, as shown in Figure 2. This contribution utilized programming concepts within CONTINUOUS to represent them as a hierarchical graph. Its objective is to facilitate learners' comprehension of the relationship between each concept in a specific programming language. Furthermore, it enables learners to identify prerequisite concepts that must be understood before engaging with their concepts of interest. The second contribution is the hints provided in programming exercises, as shown in (3) in Figure 4. This contribution also utilized programming concepts within CONTINUOUS to identify the concepts associated with each question. Through this identification process, a list of relevant concepts, referred to as hints, is displayed in each programming question to support learners in discovering solutions. Finally, the third contribution is the suggestion for the next programming concepts, as shown in Figure 12. In alignment with the identification process described in the second contribution, we observed an additional advantage beyond its use as a hint. This identification was also employed to recommend the next programming concept by exploring concepts that frequently cooccurred with those selected by learners, as demonstrated in Figure 11 and Table 1. By utilizing this frequency and learners' behavior, a list of the next programming concepts is provided as options.

All these contributions demonstrate how CONTINUOUS could be utilized in a programming support system. Moreover, they are related to learners' behavior, since learners select programming concepts, practice programming exercises, and then receive the next programming concepts. The questionnaire results in Table 3 highlight learners' perceptions of the system, examining five main features, three of which relate to the adoption of CONTINUOUS. The average rating for "technology acceptance" and "satisfaction" were above the neutral point, indicating that participants were satisfied with using the system to practice programming. On the other hand, the average rating for "mental load" and "mental effort" were close to neutral, suggesting that the system did not place excessive cognitive strain on learners. These findings indicate that the system can support learners in practicing programming with little cognitive load.

Regarding the development of the adaptive mechanism in this study, we reviewed prior research on personalized and adaptive systems. Although some studies focus on adapting learning materials (e.g., Klašnja-Milićević et al., 2011; Chookaew et al., 2014), programming education emphasizes problem-solving, making frequent practice essential (Amoako et al., 2013; Mbunge et al., 2021; Zhang et al., 2013). Accordingly, several studies have proposed systems that adapt programming exercises to learner performance—such as using pre-test scores and activity scores (Fabic et al., 2018) or fuzzy weighting and the Revised Bloom's Taxonomy (Troussas et al., 2021). Although there are various techniques in these adaptive strategies, these findings suggest that a skill-based adaptation of programming assignments can improve learners' performance and support skill development.

In our prior work, we introduced an adaptive learning support system based on an ontology of multiple programming languages (Na Nongkhai et al., 2023). This previous study involved representing an initial ontology-based adoption and designing adaptive learning strategies by utilizing the ERS in an educational setting. Building on this foundation and the benefits of adaptive learning, we successfully adopted the ERS to an ontology-based adaptive learning system. This integration addressed the second research question: "How can the Elo Rating System in an educational setting be adopted in an ontology-based adaptive learning system?" For this adoption, Formulations (1) and (2) from the ERS were utilized to update learners' skills and exercise difficulty following the successful evaluation of a code submission. When a learner submits a correct answer, their skills increases, while the exercise difficulty of that question decreases, as shown in Figure 13. After that, the system selects an exercise whose difficulty closely matches the current learner's skill, as shown in Figure 14, ensuring that each learner received exercises appropriate to their ability. Moreover, this study also defines the learning rate, representing by parameter K in Formulations (1) and (2). Variations in K affect the slope of skill increments: a low K value results in minimal skill increments, while a high K value leads to substantial gains, as shown in Figure 15. To ensure balanced progression, the parameter K is determined based on the number of questions across the difficulty levels and the concept selection, as shown in Table 2.

Furthermore, the learning logs and the learning perception responses of engagement in ADVENTURE address the third research question: "What disparities exist in learning performance, including learning achievement and perception, between learners using the adaptive support mode of ADVENTURE and those following a random practice order?" Regarding learning perceptions, the questionnaire results show that the control group (participants engaged in a random practice order) experienced more effort and pressure than the preliminary and experimental groups (participants engaged in ADVENTURE's adaptive support modes), with the highest average rating. Additionally, the control group reported a lower satisfaction with the system compared to the experimental group, as shown in Table 3. To explore more details about this, we analyzed the learning log data, including code submissions and success rates, focusing on the experimental group and the control group. Both groups consisted of students in the same academic year under the same instructor's guidance. The results indicate that the experimental group achieved significantly higher success rates in the submission of correct answers, as shown in (1) in Figure 17, compared to the control group. Additionally, the experimental group also achieved higher success rates in programming concepts, as shown in (1) in Figure 18, compared to the control group, and finally the experimental group exhibited fewer unsuccessful programming concepts than the control group, as shown in (2) in Figure 18. These differences suggest that the adaptive support mode enables learners to practice programming skills more effectively. In contrast, the control group faced challenges due to the random order of exercises, often leading them to abandon tasks perceived as too difficult. Moreover, feedback from the control group suggests a preference for a structured progression to exercises, starting with simpler tasks and gradually increasing in difficulty.

With regard to the implications for practice, the findings in this study suggest that the adaptive mode in ADVENTURE effectively supports programming practice. Instructors can apply the system to enhance instructional strategies in classroom settings, while educators and designers may adopt its mechanisms to promote personalized learning. Moreover, this study also contributes theoretically by extending the application of ontologies in education and reinforcing the link between knowledge representation and adaptive learning support.

### 7.2. Conclusions

In conclusion, these findings demonstrate that ADVENTURE facilitates programming skills' development, whether through the integration of CONTINUOUS or the use of a real-time skill assessment by ERS in an educational setting. The system enables learners to revisit and reinforce previously studied concepts (based on the list of never tried and incomplete questions in Figure 10), enhances programming logic through practical exercises, and provides a user-friendly interface that encourages continued engagement. These outcomes suggest that ADVENTURE can serve as a valuable tool for promoting personalized and adaptive programming education.

### 7.3. Limitation

The limitation in this research was that the usage of ADVENTURE was limited, because it was hosted on a university server with IP address restrictions, meaning learners could access ADVENTURE only within the university area. The participants were utilizing our system with the support of the instructors at their lectures. Moreover, participants were encouraged to use our system outside the classroom when they had free time at the university. However, the participants could not use our system when they were outside the university. This limitation caused some participants to experience technical issues while utilizing ADVENTURE, as one noted, "Frequent disconnections, server delays". According to this limitation, we could only select participants from two classes that was taught by the same instructor. This arrangement ensured timely support in the event of technical difficulties and facilitated a division into experimental and control groups. Moreover, this participant selection process led to limitations on the number of participants in the experiment, as detailed in the conclusion of Section 5, resulting in a small sample size in each group. This limited sample size presented challenges in achieving a balanced distribution concerning gender and representation across the different groups. This limitation may restrict the generalizability of the findings. In future studies, it is recommended that a larger sample size be employed to obtain more comprehensive quantitative results.

### 7.4. Future Work

In future work, we plan to expand the number of participants for experimentation to obtain balanced groups, in contrast to the experiment limitations that were described in Section 5. Moreover, we plan to improve some features in our system according to the participant feedback. For example, adding more explanation in the feedback to code submissions, add more explanations to the next programming concept suggestion, or adding more detail in the hints to each question. Furthermore, we plan to compare an adaptive approach in ADVENTURE with scenarios other than random exercises, such as situations where learners practice programming in order of difficulty without an adaptive mechanism.